\documentclass[onecolumn,reprint,showpacs,superscriptaddress,
footinbib,citeautoscript]{revtex4-1}

\usepackage{graphicx}

\usepackage{amssymb}
\usepackage{amsmath}
\usepackage{amsfonts}
\usepackage{mathrsfs}
\usepackage{lipsum}

\usepackage{color}

\usepackage[colorlinks=true,citecolor=blue]{hyperref}
\hypersetup{colorlinks=red,citecolor=blue,linkcolor=red
,urlcolor=red}

\begin{document}

\title{ Strongly anisotropic RKKY interaction in monolayer black phosphorus  }

\author{Moslem Zare}
\affiliation {School of Physics, Institute for Research in Fundamental Sciences (IPM), Tehran 19395-5531, Iran}
\author{Fariborz Parhizgar}
\affiliation{School of Physics, Institute for Research in Fundamental Sciences (IPM), Tehran 19395-5531, Iran}
\author{Reza Asgari}
\affiliation{School of Physics, Institute for Research in Fundamental Sciences (IPM), Tehran 19395-5531, Iran}
\affiliation{School of Nano Science, Institute for Research in Fundamental Sciences (IPM), Tehran 19395-5531, Iran}

\date{\today}

\begin{abstract}
We theoretically study the Ruderman-Kittel-Kasuya-Yosida (RKKY) interaction in two-dimensional black phosphorus, phosphorene. The RKKY interaction enhances significantly for the low levels of hole doping
 owing to the nearly valence flat band. Remarkably, for the hole-doped phosphorene, the highest RKKY interaction occurs when two impurities are located along the zigzag direction and it tends to a minimum value with changing the direction from the zigzag to the armchair direction. We show that the interaction is highly anisotropic and the magnetic ground-state of two magnetic adatoms can be tuned by changing the rotational configuration of impurities. Owing to the anisotropic band dispersion, the oscillatory behavior with respect to the angle of the rotation is well-described by $\sin(2k_{\rm F}R)$, where the Fermi wavelength $k_{\rm F}$ changes in different directions. We also find that the tail of the RKKY oscillations falls off as $1/R^2$ at large distances.

\end{abstract}
\maketitle

\section{Introduction}

Recently, a new type of material consisting of a black phosphorus monolayer, called phosphorene, has emerged as a promising material in the field of two-dimensional (2D) systems~\cite{Li14, Liu14, andre}. Unlike graphene which has a planar honeycomb lattice, phosphorene forms a puckered honeycomb structure due to the $sp^3$ hybridization~\cite{Liu15, Das14, Kamalakar15, Lu14, Fei14, Wang1, zare-bp}. The mobility of the charge carriers in phosphorene is quite high at room temperature $\sim 1000$ cm$^2$ V$^ {-1} $ s$^{-1}$~\cite{Li14, Liu14} which exhibits a strongly anisotropic behavior in the phosphorene-based field effect transistor with a high on/off ratio, $\sim 10^4$~\cite{Liu14}. Moreover, the band structure of black phosphorus has a direct gap at the $\Gamma$-point, can be varied by changing the number of layers and applying strain, spanning a wide range in the visible light spectrum from $0.8$ to $2$ eV~\cite{Liu14, Du10, Tran14, cakir14, Liang14, Rodin14}. These properties make phosphorene a quite interesting 2D material to be used in the field of electronics together with optoelectronics. Besides, an ambitious goal for utilizing phosphorene may be referred to next generation of spintronic devices based on the spin degrees of freedom which is almost dissipationless unlike those based on the charge degrees of freedom~\cite{fabian, Babar}.

Since pristine phosphorene is a nonmagnetic semiconductor, it cannot be directly used in the field of spintronics, however, different strategies such as edge cutting structures~\cite{Zhu14,Du15,Xu16,Farooq15,Farooq16}, doping with non-magnetic adatoms \cite{Zheng15, Yang16, khan15}, or atomic defects (vacancies and adatom adsorption) \cite{Srivastava15, Suvansinpan16, Babar} have been suggested to induce magnetic behaviors in phosphorene.
For instance, it is found that due to the magnetic instability induced by the half-filled edge bands crossing the Fermi level, phosphorene nanoribbons (PNRs) with bare zigzag edges are antiferromagnetic semiconductors \cite{Du15}. This magnetization also causes to the opening of a significant direct band gap (about 0.7 eV),
which transforms the metallic PNRs into semiconductors ones. A tensile strain about $\%5$, can destroy this antiferromagnetic ground state and shifts the semiconducting PNR into a non-magnetic metallic one.

One of the most important proposals to make a nonmagnetic material applicable in the field of spintronics is to dope phosphorene with $3d$ transition metal adatoms. Adsorption or substitutional doping of $3d$ transition-metal atoms on phosphorene \cite{Zhai17, Khan16, Kulish, Seixas15, Sui15} is thought to be the most effective approach to induce magnetization with large Curie temperature.
Since the adsorption energy of the transition-metal atoms of phosphorene is much larger than materials such as graphene, these magnetic atoms interact much stronger with phosphorene and as a result, it becomes more probable to turn the pristine phosphorene into a ferromagnetic or antiferromagnetic material due to the transfer of spin-polarized electrons from transition metal adatoms to the phosphorene \cite{Kulish}. The magnitude of the created magnetic moment depends on the metal species and the result can be tuned by the applied strain~\cite{Hu15, Sui15, Seixas15, Yu16, Cai17}.
Based on the calculations by Cai et. al \cite{Cai17}, both magnetic moments and the magnetic ordering of $3d$ transition metal atoms adsorbed on phosphorene system can be tuned by strain. Recently, Seixas et. al \cite{Seixas15} reported that the magnetic coupling can be engineered by a potential gate in the phosphorene monolayer doped with Co atoms, and a ferromagnetic to antiferromagnetic phase transition is observed.

Ruderman-Kittel-Kasuya-Yosida (RKKY)~\cite{Kasuya, Ruderman} interaction is an indirect exchange interaction between two magnetic impurities mediated by a background of conduction electrons of the host material and is the most important mechanism of the coupling between magnetic impurity dopants in metals and semiconductors. As an applied point of view, this interaction can lead a magnetically doped system to interesting phases such as ferromagnetic \cite{Vozmediano, Brey, Priour, Matsukura, Ko, Ohno-science}, antiferromagnetic \cite{Minamitani, Loss15}, spiral \cite{moslem-si, Mahroo-RKKY} and spin-glass \cite{pesin, Eggenkamp, Liu87}. Besides the practical importance of the RKKY interaction in the possible magnetic phases of semiconductor, it can provide information about the intrinsic properties of the material since this interaction is proportional to the susceptibility of the host system.
While this interaction falls off by $R^{-D}$, where $D$ is the dimension of the system \cite{fariborz-blg, fariborz-mos2}, it oscillates with the Fermi wavevector originates from the Friedel oscillations. In systems with multiband structure \cite{fariborz-mos2} or with spin polarization \cite{fariborz-sp} these oscillations become more complicated than a monotonic oscillation with $\sin(2k_{\rm F}R)$ behavior where $k_{\rm F}$ is the wave vector of the electrons (holes) at the
Fermi level. Moreover, it has been shown that the magnitude of the RKKY interaction can be severely affected by the density of states (DOS) at the Fermi energy \cite{fariborz-blg, Mahroo-RKKY}. In addition, it can be sensitive to the direction of the distance vector between impurities in materials such as graphene \cite{sherafati-g, fariborz-blg} owing to the bipartite nature of the honeycomb sublattice. In materials with Rashba spin-orbit coupling, the spin response of the system to the magnetic impurity depends on the direction of the magnetic moment \cite{Mahroo-single} and as the result the RKKY interaction becomes anisotropic \cite{Mahroo-RKKY}.

In the present work, we use the Green's function technique in order to derive the RKKY interaction in a monolayer phosphorene and study the effect of its anisotropic band dispersion on the magnetic interaction between impurities. Most studies of the RKKY in different system, have been mainly based on the isotropic Fermi surfaces, although there are a few studies on anisotropic samples in III-V semiconductors \cite{pesin, Timm, Patrone, Davison, Han12}. The highly anisotropic band structure of phosphorus gives rise to a highly anisotropic indirect interaction and it would be worth to survey this effect on phosphorene. The RKKY interaction is dependent on the direction that separates the impurities, and on band filling, with nearly flat structure in the valence band playing an important role in enhancing the impurity coupling.

This paper is organized as follows. In Sec. \ref{sec:theory}, we introduce a model Hamiltonian and then establish the theoretical framework which is aimed to calculate the indirect exchange interaction. The charge excitations in phosphorene may also be described by a simple model Hamiltonian for which we derive an analytic expression of the RKKY interaction.  In Sec. \ref{sec:results}, we present and describe our numerical results for the proposed magnetic doped phosphorene. Finally, our conclusions are summarized in Sec. \ref{sec:summary}.

\section{Theory and Method}\label{sec:theory}
\subsection{Model Hamiltonian}

\begin{figure}
 \includegraphics[width=0.9\linewidth]{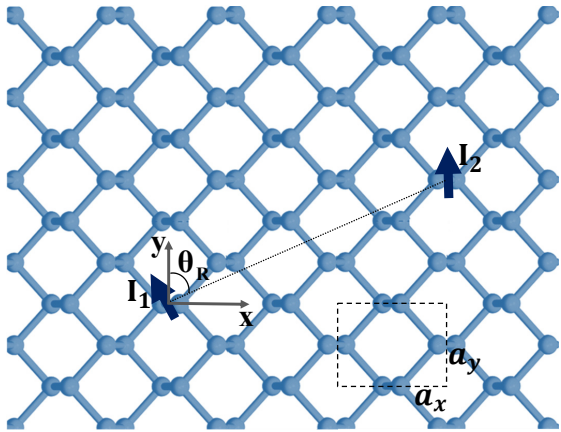}
 \caption{(Color online) A top view schematic of the phosphorene honeycomb-like lattice a with rectangle primitive cell with lattice constants $a_x$ and $a_y$.
 Two magnetic impurities and azimuthal angle between them denote by ${\bm I_1, \bm I_2}$ (navy arrows) and $\theta_{\bf R}$, respectively.
\label{fig:lattice}}
 \end{figure}

Black phosphorus has an orthorhombic crystal structure composed of quasi two-dimensional sheets, where each sheet known as phosphorene, attached together through a weak inter-layer Van der Waals interaction. Figure \ref{fig:lattice} shows the lattice structure of phosphorene, where the unit cell is shown by black dashed lines. This unit cell consists of four atoms and the lattice constants along the $x$ (armchair) and $y$ (zigzag) directions are respectively $a_x= 4.63$\AA \,and $a_y = 3.3 $\AA ~\cite{Rodin14}.
The two-dimensional $\bm{k}\cdot\bm{p}$ model Hamiltonian ${\cal H}_0$ of phosphorene  can be written as~\cite{Rodin14,zare-bp,asgari15}

\begin{equation}\label{H0}
{\cal H}_{0}=
\begin{pmatrix}
E_c+\eta_c k_x^2+\nu_c k_y^2 &\gamma k_x& \\ \gamma k_x& E_v-\eta_v k_x^2-\nu_v k_y^2
 \end{pmatrix},
\end{equation}
in the basis of the conduction and the valence bands spinors $(\Psi_{c},\Psi_{v})$, respectively. $E_c$ and $E_v$ are the energies of the conduction and valence band edges respectively given by $E_c=-0.3797$ eV and $E_v = -1.2912$ eV. Other coefficients are $\eta_c=0.008187$, $\eta_v=0.038068$, $\nu_c= 0.030726$, $\nu_v= 0.004849$ which are in units of eV nm$^2$, and $\gamma= 0.48$ eV nm.
The energy dispersion of the Hamiltonian is obtained like as
\begin{eqnarray}
E_{\tau}=\frac{1}{2}[H_c + H_v + \tau\sqrt{4H_{cv}^{2}+(H_{c}-H_{v})^{2}}]\label{disp},
\end{eqnarray}
where $H_{c}=E_{c}+\eta_{c}k_{x}^{2}+\nu_{c}k_{y}^{2}$, $H_{v}=E_{v}-\eta_{v}k_{x}^{2}-\nu_{v}k_{y}^{2}$, $H_{cv}=\gamma k_{x}$, and $\tau=\pm 1$ denotes the conduction or valence band. The Hamiltonian \eqref{H0} resembles two elliptical Hamiltonian $H_c$ and $H_v$ which are connected through a coupling term $H_{cv}=\gamma k_x$. Therefore,  in the limit of zero $\gamma$, the problem of phosphorene reduces to two anisotropic 2D electron gas systems.

\begin{figure}[]
\begin{center}
\includegraphics[width=3.4in]{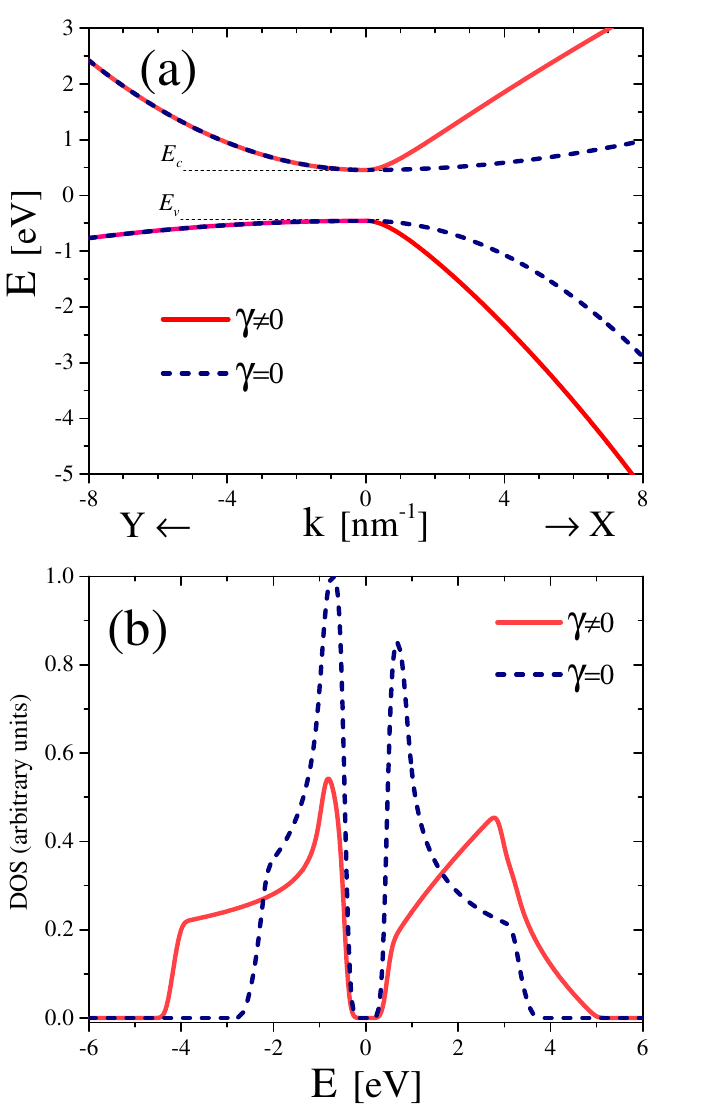}
\end{center}
\caption{\label{Fig:dis} (Color online) (a) The dispersion relation of monolayer phosphorene along the $Y-\Gamma-X$ direction in the Brillouin zone. Note that the zero-point energy is shifted to the middle of the energy band gap. (b) DOS (in an arbitrary units) of monolayer phosphorene deduced from the $\bm{k}\cdot\bm{p}$ model Hamiltonian Eq. \ref{H0}. For the sake of competence, the impact of the interband coupling $\gamma$ is also explored, therefore, the band dispersion and DOS in the case of $\gamma=0$ are plotted (dashed curves).}
\end{figure}

Figure \ref{Fig:dis}(a) shows the anisotropic band structure of the monolayer phosphorene along the $Y-\Gamma-X$ direction in the Brillouin zone where $\Gamma$ is the $k=0$ point and $X, Y$ show the edges of the Brillouin zone in the $x$ and $y$ directions of the momentum space, respectively. It should be noted that throughout the paper the zero-point energy is shifted to the middle of the band gap.

Figure~\ref{Fig:dis}(b) shows the DOS of phosphorene in both $\gamma \neq 0$ and $\gamma=0$ limits.
Note that, since the integration over ${\bf k}$ is restricted to the first Brillouin zone region, the band width both in the conduction and valence bands are limited.
 Distinguished electron-hole asymmetry of phosphorene from the conventional 2D semiconductors is unique where the nearly flat band of the spectrum for holes along the $y$-direction (see Fig.~\ref{Fig:dis} (a)) results in a high value of the DOS in the valence band near the gap edge. The existence of the Van Hove singularity in phosphorene has previously addressed~\cite{Huang15}.
Owing to the large effective mass of the holes along the $\Gamma-Y$ direction, the RKKY coupling may significantly intensify.
In agreement with our result, once the Fermi energy crosses the edge of the gap where the DOS is large, a ferromagnetic ~\cite{Fleck, Hlubina} and an antiferromagnetic~\cite{Hirsch} features of the RKKY interaction can be significantly enhanced.

\begin{figure}
\centering
\includegraphics[width=3.5in]{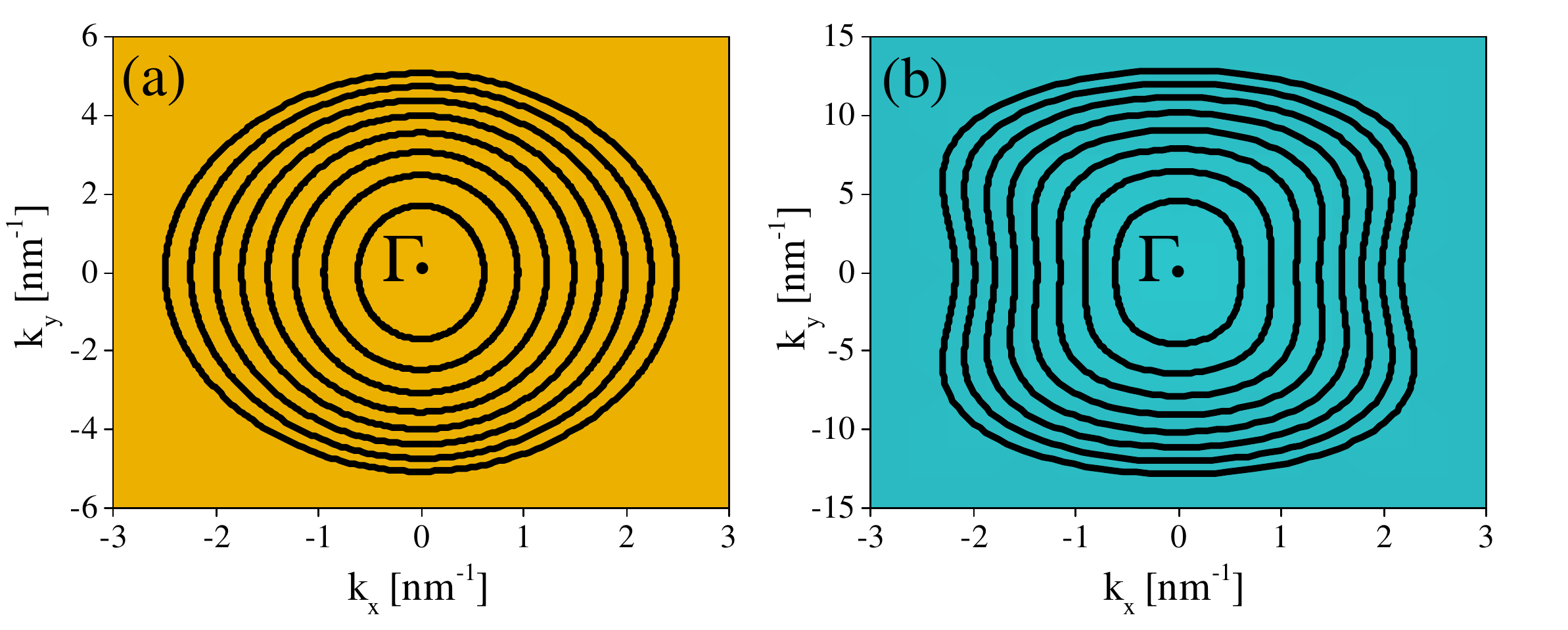}
\caption{(Color online) Iso-energy contour plot in the $(k_x,k_y)$ space, respectively, for the conduction (a) and valence bands (b). The energy values of the black lines starts (interior band) from $E = 0.5$eV ($E=-0.5$eV)in the electron (hole) doped case and its value increases with step of $0.1$ eV.}
\label{contours}
\end{figure}

Figure \ref{contours} shows the iso-energy lines in the $(k_x,k_y)$ plane. The iso-energy lines indicate energies $\pm 0.5$eV in the interior line and its absolute value increases with the step of $0.1$eV for both the conduction (a) and valence (b) bands.
As seen, while the bands resemble anisotropic elliptical lines in the conduction, the band structure posses more complicated behavior in the valence bans specially along the $k_x$ direction.

\subsection{RKKY Interaction}\label{sec:Formalism}

Our system incorporates two localized magnetic moments
whose interaction is mediated through a phosphorene charge carriers. The contact interaction between
the spin of itinerant electrons and two magnetic impurities is $V=\sum_j v_{j}$. The interaction term between each impurity and the electrons of the host material is $v_j=-j_c {\bm I_j}\cdot{\bm s}\delta({\bm r}-{\bm r}_j)$ where ${\bm s}$ is the spin density operator of the conduction electrons of the host and ${\bm r_j}$ is the position vector of the $j$th magnetic adatom. Moreover, $I_j$ denotes the $j$th impurity's spinor and $j_c$ is its coupling strength with the host material. In the linear response theory the energy exchange interaction between the two magnetic moments ${\bm I_1}$ and ${\bm I_2}$, mediated by the internat charge carriers, obtains from the expectation value $E({\bf r}_i- {\bf r}_j) = <\Psi|V({\bf r}_i- {\bf r}_j)|\Psi>$, where $\Psi$ is the perturbed wavefunctions due to the presence of the one magnetic impurity and ${\bf r}_i- {\bf r}_j$ is the actual relatively position of the first and second magnetic atoms.
Using the second order perturbation theory~\cite{Ruderman, Kasuya, Imamura}, the potential Hamiltonian according to two of impurities located respectively at ${\bf r}_i, {\bf r}_j$, is given by

\begin{eqnarray}
V_{ij}({\bf r}_i- {\bf r}_j) =  J({\bf r}_i, {\bf r}_j){\bm I_i}\cdot{\bm I_j},
\end{eqnarray}
where $J({\bf r}_i, {\bf r}_j)=(j_c^2\hbar^2/4)\chi({\bf r}_i- {\bf r}_j)$ is the exchange integral and $\chi({\bf r}_i- {\bf r}_j)$ is the spin susceptibility of the itinerant charge carriers and determines the
indirect interaction between two local moments. The spin susceptibility can be obtained from the retarded Green's function as

\begin{eqnarray} \label{eq:RKKY}
\chi({\bf r}_i,{\bf r}_j) &=& -\frac{2}{\pi} \Im m\int_{-\infty} ^{\varepsilon _{\rm F}} dE
\nonumber\\
&&Tr[G_0({\bf r}_i,{\bf r}_j;E +i\epsilon)G_0({\bf r}_j,{\bf r}_i;E +i\epsilon)],\nonumber\\
\end{eqnarray}
where $G_0({\bf r}_i,{\bf r}_j;E)$ is the noninteracting Green's function. In this equation, the factor 2 accounts the spin degeneracy, $\varepsilon_{\rm F}$ is the Fermi energy, trace should be taken over the degrees of freedom and $\epsilon$ is an infinitesimal quantity guarantees the Green's function to be retarded. In the linear response regime where one neglects the three particle interaction, the total Hamiltonian of the impurities can be obtained as a superposition of all $V_{ij}$s.
The corresponding matrix element of the real-space Green's function is obtained from the Fourier transformations of $G_0(\bm k, E)$ element as

\begin{align}\label{eq:a0}
G_0({\bf r}_i,{\bf r}_j,E)=\frac{1}{\Omega_{BZ}}\int  d{\bf k} e^{i{{\bf k}\cdot({\bf r_i}-{\bf r_j})}}G_0({\bf k},E),
\end{align}
where $\Omega_{\text{BZ}}=(2\pi)^2/a_x a_y$ is the area of the first Brillouin zone and the $k$ space Green's function, $G_0({\bm k},E)=(E-H+i\epsilon)^{-1}$ is given by
\begin{equation}\label{gg}
G_0(k,E)=
[(E-H_c)(E-H_v)-H_{cv}^2]^{-1}
\begin{pmatrix}
 E-H_v& H_{cv}\\H_{cv}&E-H_c\\
 \end{pmatrix}.
\end{equation}

Notice that it would be difficult to find an analytic expression of the RKKY interaction in general case based on Eq. \eqref{H0}. However, we derive an analytic expression of the RKKY in a simple case where $\gamma=0$ and afterwards, we will present our
numerical results using Eq. \ref{eq:RKKY}.

\subsection{Analytic solution for the case that $\gamma=0 $}
Since in the absence of the coupling term $\gamma$ between conduction and valence bands, the spectrum of phosphorene reduces to two separate anisotropic ovals, an important question is that can one consider phosphorene as two separate anisotropic 2D systems? In this subsection, we set $\gamma=0$ and obtain analytic solutions for two highly-interest orientations of two magnetic impurities namely along the zigzag and armchair directions. Comparison of these analytic results with the results of the RKKY interaction, using full Hamiltonian \eqref{H0}, will clarify the status of the coupling parameter, $\gamma$. We first obtain the Green's functions in the real space using Eq.~\ref{gg} and then calculate the susceptibility in the real space and equivalently the RKKY interaction.

Let us consider two impurities where one impurity is located at the origin and the other remains in a position ${\bf R} $. In the limit of $\gamma=0$, the real space Green's function can be obtained by

\begin{align}\label{eq:a1}
G_0({\bf R},{0},E)=\frac{1}{\Omega_{BZ}}\int  d{\bf k} e^{i{\bf k.R}}
 \begin{pmatrix}
G_0^{c}({\bm k},E)& 0 \\ 0&G_0^{v}({\bm k},E)\\
 \end{pmatrix},
\end{align}
where $G_0^{s}=(E-E_s-\eta_s k_x^2-\nu_s k_y^2)^{-1}$ with $ s=c$ or$v$ corresponding to the conduction and valence bands, respectively, and then the spin susceptibility can be computed by using

\begin{eqnarray}\label{eq:a6}
\chi({\bf R,0})&=&-\frac{2}{\pi}\Im m \sum_{s=c,v}\int_{-\infty}^{\varepsilon_{\rm F}}dE
 G_0^{s}{({\bf R},0,E)}G_0^{s}{(0,{\bf R},E)}. \nonumber\\
\end{eqnarray}

In a special case when both impurities are located along the armchair direction ($R_y=0$, which corresponds to the $\Gamma X$ direction in the $k$-space), we obtain the real space Green's functions and after some straightforward calculations (see Appendix) we have

\begin{align}\label{eq:F1}
 G_0^{s}{( R_x,{0},E)}= \frac{(-1)^{p_s} \pi}{2\sqrt{\eta_s\nu_s}\Omega_{BZ}}K_0(m_s\sqrt{a_s}),
\end{align}
where $a_s= (-1)^{p_s} \frac{E-E_s}{\nu_s}, m_s=R_x\sqrt{\nu_s/\eta_s}$ and $K_n(x)$ is the modified Bessel function of the second kind of integer orders $n$. The factor ${p_s}=1,2$ denotes $c$ and $v$, respectively.
The same procedure can be followed along the zigzag direction (which corresponds to the $\Gamma Y$ direction in the $k$-space) and $G^s_0(R_y,0)$ will be achieved by replacing $(R_x,\eta_{s}, \nu_{s})$ with $(R_y,\nu_{s}, \eta_{s})$ respectively in the above equation.

Using obtained Green's functions, we can calculate the spin susceptibility $\chi$ by evaluating the integral in Eq. \ref{eq:a6} over energy. We discuss this integration for the electron (n-) and hole (p-) doped cases, separately.
First, we evaluate the RKKY interaction $\chi(R_x,0)=\sum_{s=c,v} \chi^s(R_x,0)$ for the n-doped phosphorene using the Eq.~\eqref{eq:a6} as

\begin{eqnarray}\label{eq:d1}
\chi^{s}{(R_x,0)}=\frac{-2}{\pi}\Im m\int_{-\infty}^{\varepsilon_{\rm F}}dE
 G_0^{s}{(R_x,0,E)}G_0^{s}{(0,R_x,E)}.\nonumber\\
\end{eqnarray}

For the sake of simplicity, we define $ K_0(m_v\sqrt{a_v})=K_0(\sqrt{u+i\epsilon^{+}})$ where $u=m_v^2(E-E_v) / \nu_v$, in which $E=\varepsilon+i\epsilon^{+}$ then for a positive number $u$, $K_0(\sqrt{u+i\epsilon^+})=K_0(\sqrt{u})$ and $K_0(\sqrt{-u \pm i\epsilon^+})=(-\pi/2)[Y_0(\sqrt{u})\pm iJ_0(\sqrt{u})]$. In order to carry out the integration Eq.~\eqref{eq:d1}, we decompose the integral into two parts, $\int_{-\infty}^{\varepsilon_{\rm F}}=\int_{-\infty}^{E_v}+\int_{E_v}^{\varepsilon_{\rm F}}=t_0+t_1$, where the first term accounts

\begin{eqnarray}\label{eq:d2}
t_0\varpropto \int_{0}^{\infty} Y_0(\sqrt{u})J_0(\sqrt{u})du.
\end{eqnarray}

Following the approach introduced in \cite{Saremi}, we multiply the integrand by different cutoff functions namely;
$f (u,u_0)=\exp(-u/u_0)$ and $f (u,u_0)=\exp(-u^2/u_0^2)$ and then take the limit $u_0\rightarrow\infty$. In this limit, the cutoff function $f (u,u_0)\rightarrow 1$ and the contribution of the integral $t_0$ become nearly zero.

Because $\Im m K_0(\sqrt{u})=0$ for $E>E_v$, thus we have $t_1=0$.
This means that for the n-doped phosphorene, the contribution of the $\chi^{v}{( 0,R_x)}$ in the RKKY interaction is zero, however, the contribution of the $\chi^{c}{(R_x,0)}$ is finite and given by
\begin{eqnarray}\label{eq:d4}
\chi^{c}{(R_x,0)}=\frac{-\pi}{{2\eta_c \nu_c \Omega_{BZ}^2}}\Im m\int_{-\infty}^{\varepsilon_{\rm F}}dE K^2_0(m_c\sqrt{a_c}).
\end{eqnarray}

Here, similar to the  $\chi^{v}{(R_x,0)}$, we decompose the above integral into two terms, $\int_{-\infty}^{\varepsilon_{\rm F}}=\int_{-\infty}^{E_c}+\int_{E_c}^{\varepsilon_{\rm F}}$. Let us write $K_0(m_c\sqrt{a_c})$ as $K_0(\sqrt{v+i\epsilon^+})$ where $v=m_c^2(E_c-E) / \nu_c$ for $E<E_c$. It is easy to verify that $\Im mK_0(m_c\sqrt{a_c})=0$ for $v>0$ and it implies that the first integral becomes zero. Furthermore, $\Im m K_0(m_c\sqrt{a_c})= (\pi/2)^2Y_0(\sqrt{v})J_0(\sqrt{v})$ in the interval of the second integral $E>E_c$, and therefore we have

\begin{eqnarray}\label{eq:d5}
\chi^{c}{(R_x,0)}&=\frac{\pi^3}{{4m_c^2\eta_c \Omega_{BZ}^2}} \int_{0}^{\tilde{x}>0}Y_0(\sqrt{v})J_0(\sqrt{v})dv,
\end{eqnarray}
where $\tilde{x}=R_x^2(\varepsilon_{\rm F}-E_c)/\eta_c$. By using the following integral

\begin{eqnarray}\label{eq:d7}
\int_{0}^{\tilde{x}>0}Y_0(\sqrt{v})J_0(\sqrt{v})dv=-\pi^{-1/2}G^{2,0}_{1,3}(\tilde{x}|^{3/2}_{1,1,0}),
\end{eqnarray}
where G is Meijer-G function \cite{Meijer}, we finally derive the spin susceptibility of the n-doped phosphorene as

\begin{eqnarray}\label{eq:g01}
\chi{({R_x,0})}=\frac{-\pi^{5/2}}{{4R_x^2\nu_c \Omega_{BZ}^2}} G^{2,0}_{1,3}(\tilde{x}|^{3/2}_{1,1,0}).
\end{eqnarray}

In the case of the zigzag direction, the calculations are the same as $x$-direction and one needs to exchange $\eta_{c(v)}$ with $\nu_{c(v)}$ and $R_y$ with $R_x$, respectively. Therefore, the spin susceptibility along the $y$ direction is given by

\begin{eqnarray}\label{eq:g02}
\chi{({ 0,R_y})}=\frac{-\pi^{5/2}}{{4R_y^2\eta_c \Omega_{BZ}^2}} G^{2,0}_{1,3}(\tilde{y}|^{3/2}_{1,1,0}),
\end{eqnarray}
where $\tilde{y}=R_y^2(\varepsilon_{\rm F}-E_c)/\nu_c$. Importantly enough, by comparing Eq. \eqref{eq:g01} and \eqref{eq:g02} one can find that the RKKY interaction in n-doped phosphorene is almost four times ($\nu_c/\eta_c=3.75$) grater than that along the $x$-direction.

In the p-doped phosphorene case, after some simple and straightforward algebra, the same expression is obtained by replacing
$(E_v,\eta_{v}, \nu_{v})$ with $(E_c,\eta_{c}, \nu_{c})$ and $\tilde{y}$ by $-\tilde{y}$.

In order to calculate the long-range behavior of the RKKY interaction, we do need to expand the Meijer-G functions \cite{Meijer} and use its asymptotic expansion. In the both electron and hole doped phosphorene along the armchair direction ($x$), the asymptotic term of the spin susceptibility reads as

\begin{eqnarray}\label{eq:g05}
\chi_s{(R_x\gg a_x,0)}=\frac{-\pi^{2}}{{8\eta_s \Omega_{BZ}^2}} \frac{\sin({2k_{\rm F}^sR_x)}}{R_x^2},
\end{eqnarray}
where $k_{\rm F}^s=[(-1)^{p_s}(E_s-\varepsilon_{\rm F})/\eta_s]^{-1}$ is the Fermi wave vector of the charge carriers. Notice that ${p_s}=1,2$ in the conduction (n-doped) and valence (p-doped) cases, respectively. In the case of the zigzag direction (y), same as (x)-direction, one can obtain the asymptotic expansion just by exchanging $\nu$ with $\eta$ and $R_y$ with $R_x$, respectively.

As shown in Eq.~\ref{eq:g05}, the resulting RKKY interaction is just similar to an ordinary two-dimensional metal and it decays like $R^{-2}$ at the long-distance limit.

\begin{figure}[]
\begin{center}
\includegraphics[width=3.5in]{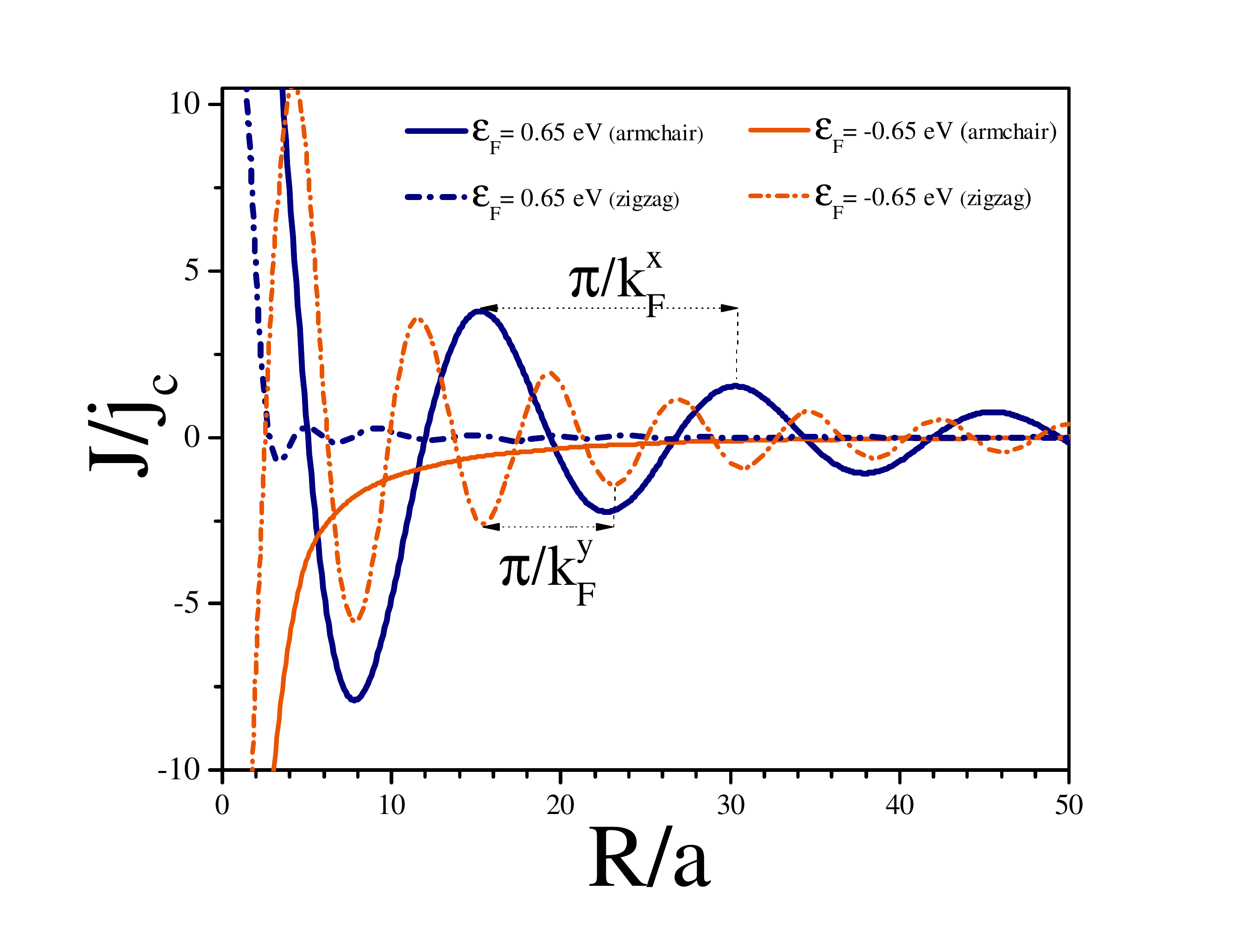}
\end{center}
\caption{ \label{Fig:3} (Color online) Short-range behavior of the analytical solutions of the RKKY interaction $J$ scaled by $j_c$ (Eqs. \eqref{eq:g01},\eqref{eq:g02}), when both impurities are located along the armchair (solid curves) and zigzag directions (dashed-dotted curves) as a function of the impurities distance for both n- (navy) and p-doped (orange) phosphorene. The wavelength of the oscillations is equal to $\pi/k_{\rm F}$ where $k_{\rm F}^{x(y)}$ is the wave vectors of the electrons and holes at the Fermi level correspond to the armchair (x) and zigzag (y) directions, respectively. Note that the distances are scaled by units of the lattice constants $a_x$ and $a_y$. Here, the Fermi energy is $\varepsilon_{\rm F}=0.65$ or $-0.65$ eV.}
\end{figure}

\begin{figure}[]
\begin{center}
\includegraphics[width=3.5in]{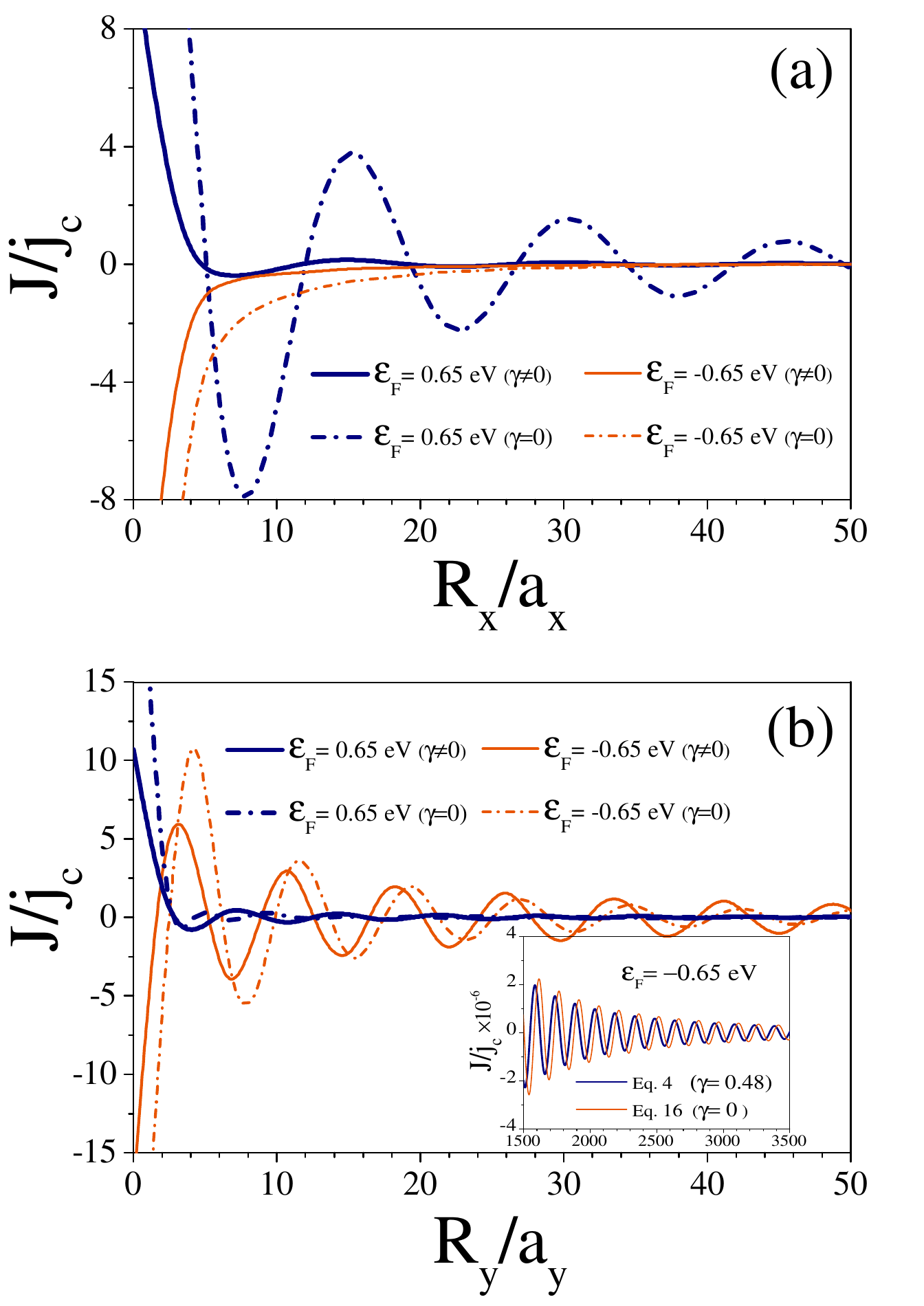}
\end{center}
\caption{ \label{Fig:4} (Color online) Short-range behavior of the numerical solutions of the RKKY interaction $J$ scaled by $j_c$, when both impurities are located along the (a) armchair and (b) zigzag directions as a function of impurities distance for both n (navy curves) and p (orange curves) doped phosphorene at finite $\gamma=0.48$ eV nm. Inset: the asymptotic behavior of the $\chi$ in doped case along the zigzag direction. This demonstrates that the tail of the RKKY oscillations falls off as $\sin({2k_{\rm F}^v R)}/R^2$ at large distances. The influence of the interband coupling $\gamma$ is also shown (dashed-dotted curves are in the case of the $\gamma=0$). Distances are scaled by units of the lattice constants $a_x$ and $a_y$, respectively. The proper results of the quasiparticle excitation are captured by considering a finite value of the $\gamma$ at given the Fermi energy. }
\end{figure}

\section{Numerical Results and discussions}\label{sec:results}

In this section, we present our numerical results of the RKKY interaction, Eq. \eqref{eq:RKKY}, and investigate its dependence on parameters such as the distance between impurities, the Fermi energy and the band structure coupling parameter $\gamma$ which in the latter case, we are able to use our analytic results Eqs. \eqref{eq:g01} and \eqref{eq:g02}.
In the numerical calculations, cutoffs are needed both on the wave vector and energy, where for the former we use $(k_{cx},k_{cy})=(2\pi/a_x,2\pi/a_y)$ and for the latter, we also apply the edge energy of the first Brillouin zone.

Suppose both the itinerant and the localized spins on the same atom, then we can write the exchange coupling in terms of the Hund’s-rule energy as $J_H$, where $j_c \hbar^2 \simeq 2J_H\sim2 eV$ \cite{Sherafati11}. Then as a tangible quantity in this paper we calculate the normalized  RKKY interaction $J/j_c$.

Figure \ref{Fig:3} shows our analytic results (Eqs. \eqref{eq:g01} and \eqref{eq:g02}) for the spacial behavior of the RKKY interaction $J$ scaled by $j_c$ as a function of the impurities distance for both n- (navy color) and p-doped (orange color) phosphorene, when impurities are located along the armchair (solid lines) and zigzag directions (dashed-dotted lines), respectively. Here, we set $\varepsilon_{\rm F}=\pm 0.65$ eV and we scale the distances along the armchair and zigzag directions to the lattice constants $a_x$ and $ a_y$, respectively. As our asymptotic formula Eq. \eqref{eq:g05} shows, this interaction have an oscillatory behavior with the period of the oscillation $2k_{{\rm F}}^s$ along $s=x$ and $y$ directions. In the hole-doped regime, the RKKY interaction illustrates higher values along the zigzag direction, while in the electron-doped regime, the $J$ becomes more important along the armchair direction. This can be understood by exploring the band dispersion, wherein the limit of $\gamma=0$ (see the Fig. \ref{Fig:dis}) the effective mass along the $x$-direction takes higher value than the $y$-direction for the conduction band, while opposite behavior occurs for the valence band.

In order to investigate the effect of the $\gamma$ term on the behavior of the RKKY interaction, we depict our numerical results of the exchange coupling as a function of $R_x$ (panel a) and $R_y$ (panel b), in Fig.~\ref{Fig:4} and compare the results with our analytical results in the case that $\gamma=0$. The $J$ displays an oscillatory behavior with respect to the distance between two impurities. In the armchair direction, as the $\gamma$ term is a coefficient to $k_x$, both the conduction and valence bands are strongly affected by this term and on the other hand, the RKKY interaction is less affected by the $\gamma$ term along the $y$-direction. Furthermore, this figure shows that our analytic results overestimate the RKKY interaction in $\gamma=0$ approximation, that could be expected from Fig. \ref{Fig:dis} (b) where the $\gamma=0$ approximation predict higher values of the density of states. The influence of the interband coupling $\gamma$ is also shown (dashed-dotted curves are in the case of the $\gamma=0$) in the figure.
 As seen in the Fig.~\ref{Fig:4}(b), by turning on the coupling parameter $\gamma$, the magnetic ground-state of the magnetic impurities can be changed from ferromagnetic to antiferromagnetic, or vice versa.

In the inset, the asymptotic behavior of the $J$ is compared with that obtained by using Eq. 16 in doped case along the zigzag direction. This demonstrates that the our numerical results of the RKKY oscillations can be faithfully fitted by an analytical expression as $\sin({2k_{\rm F}^v R)}/R^2$ for large distances between two impurities.

\begin{figure}[]
\begin{center}
\includegraphics[width=3.5in]{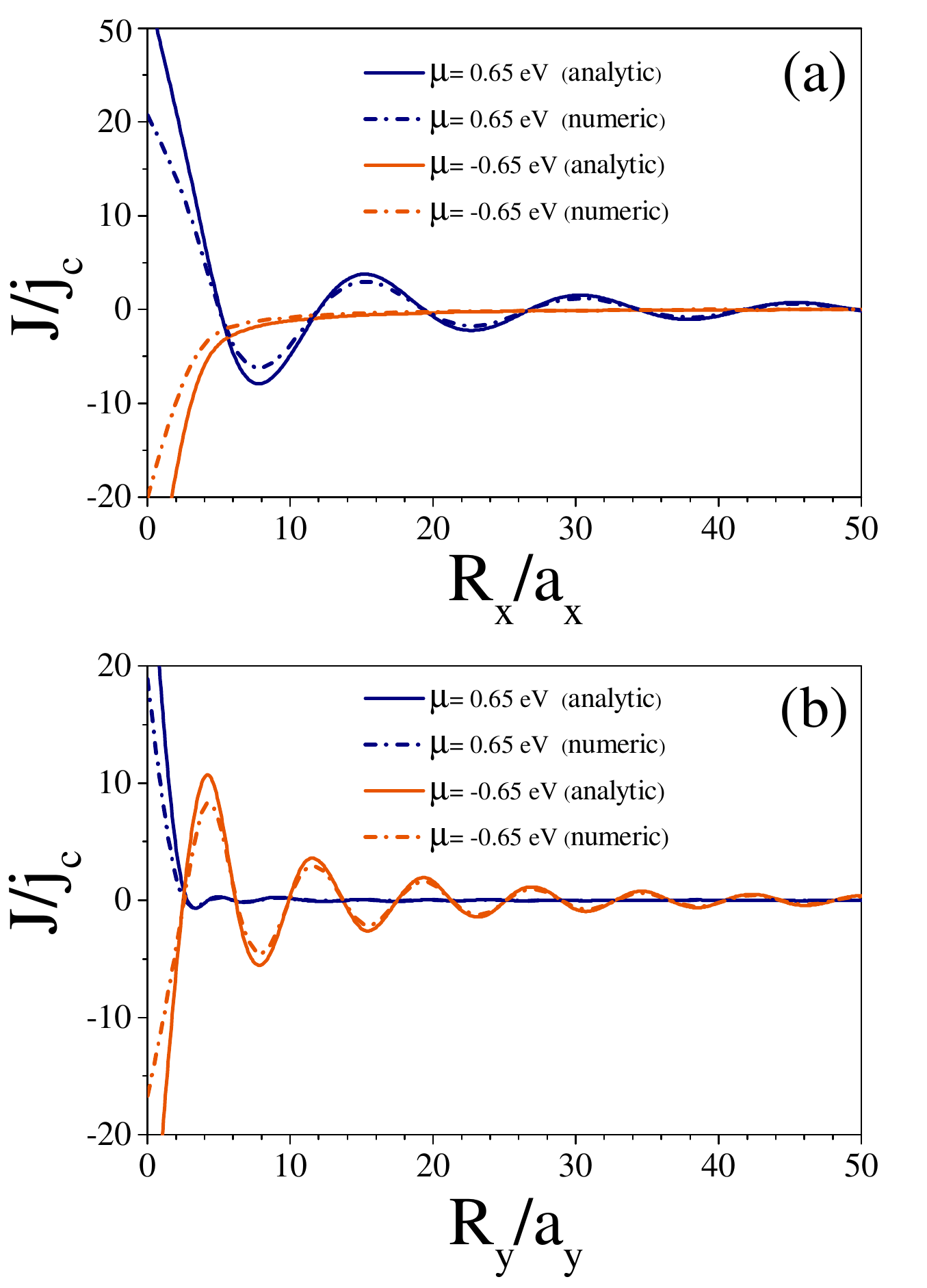}
\end{center}
\caption{ \label{Fig:new6} (Color online) Comparison of the numerical solutions of the RKKY interaction with the analytical ones, when both impurities are located along the (a) armchair and (b) zigzag directions as a function of impurities distance for both n (navy curves) and p (orange curves) doped phosphorene. Distances are scaled by units of the lattice constants $a_x$ and $a_y$, respectively.}
\end{figure}

Moreover, we compare the analytic results with those obtained by the full numerical calculations along the both armchair (a) and zigzag (b) directions. This figure is achieved in $\varepsilon_F=\pm 0.65$ for the conduction and valence bands, respectively. Our numerical results show a very good agreement of the two methods at large distances.

\begin{figure}[]
\begin{center}
\includegraphics[width=3.4in]{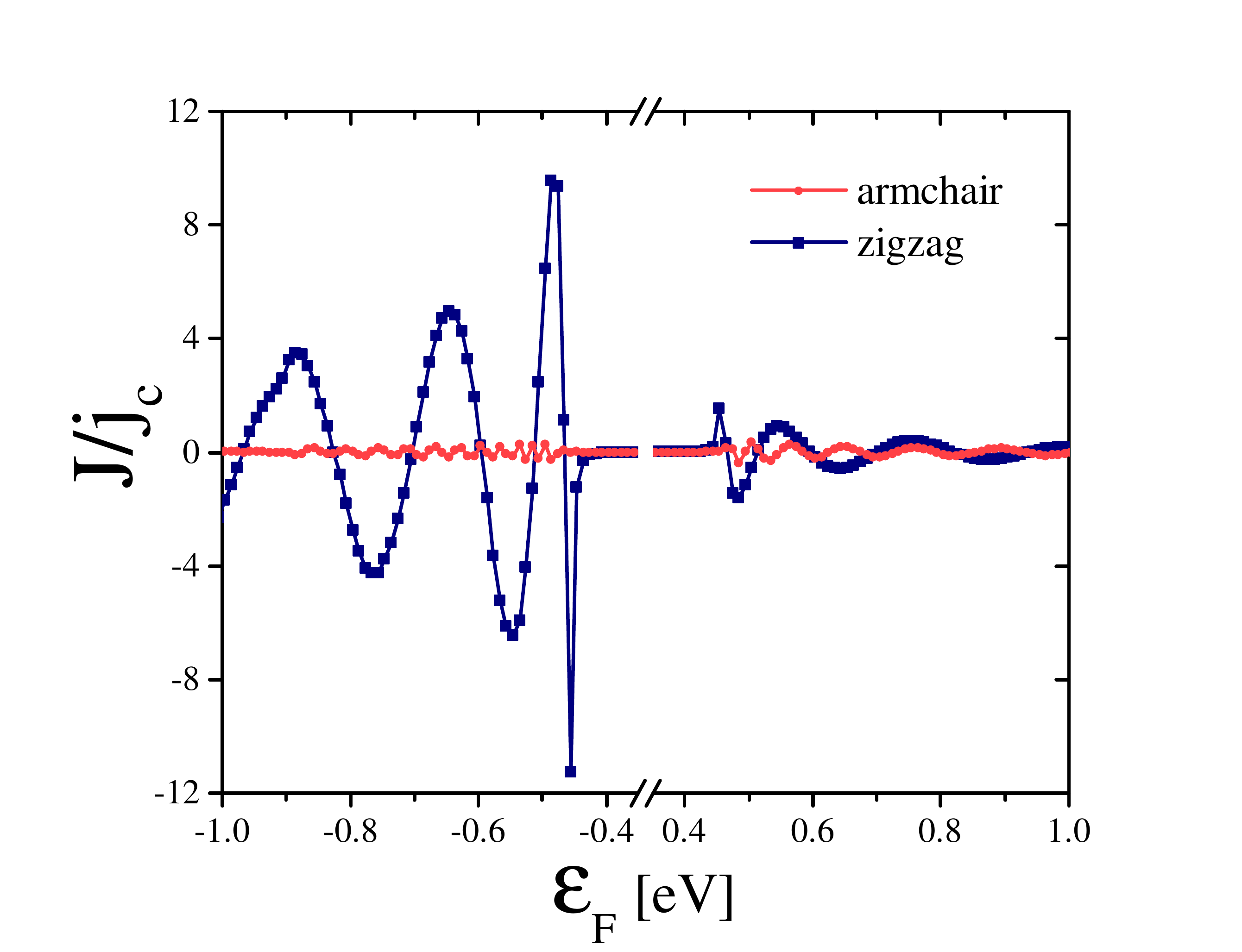}
\end{center}
\caption{\label{Fig:5} (Color online) Numerical solution of the exchange interaction $J$ scaled by $j_c$,  when both the impurities with distance $R=5$ nm located along the armchair direction (circle, red) and zigzag direction (square, navy), as a function of the Fermi energy at finite $\gamma=0.48$ eV nm. The RKKY interaction has an oscillatory behavior with the Fermi energy and the period of oscillations is different along the various directions as the band dispersion is anisotropic.
Due to the flat band at the edge of the valence band, the RKKY interaction enhances significantly. }
\end{figure}

Figure \ref{Fig:5} shows the dependence of the coupling between impurities with respect to the Fermi energy for a fixed distance between impurities $R=5$nm and two different directions namely; zigzag and armchair. As seen here,  the RKKY interaction has an oscillatory behavior with the Fermi energy. The period of oscillations is different along the various directions as the band dispersion is anisotropic, which makes $k_{{\rm F}x}\neq k_{{\rm F}y}$.

The RKKY interaction depends strongly on the shape of the Fermi surface \cite{Weismann,Lounis} which is proportional to the inverse of the multiplication of the curvatures $ \partial^2 E / \partial k_x^2$ ,$ \partial^2E /\partial k_y^2$  ($J\propto | \partial^2E / \partial k_x^2 \times \partial^2E / \partial k_y^2|^{-1}$) such that for small values of the curvature, the Fermi surface has a flat region leading to large values of the RKKY interaction. In the case that impurities located along the zigzag direction, due to the flat band of the phosphorene (see Figure \ref{contours}) curvature is strongly reduced and thus the large value of the RKKY interaction is achieved.

As the Fermi energy lies inside the gap, the RKKY interaction would be weakened, however, it still has a finite value according to the Van-Vleck mechanism. Moreover, at the edge of both the conduction and valence bands and due to the existence of the Van Hove singularities~\cite{Fleck,Hlubina,Hirsch}, the RKKY interaction enhances significantly. This means that one can expect large values of the RKKY interaction by small amount of doping.

\begin{figure}[]
\begin{center}
\includegraphics[width=3.4in]{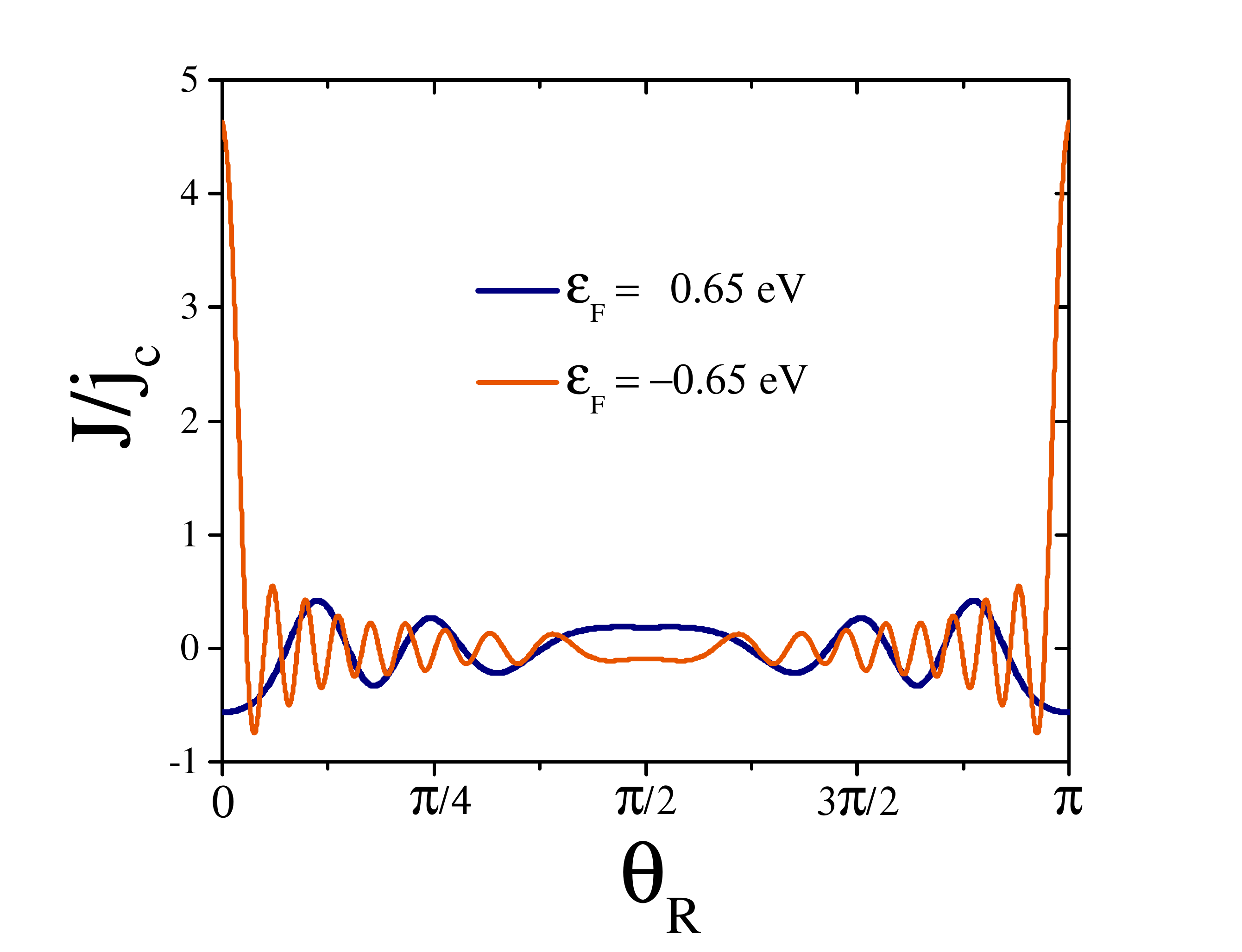}
\end{center}
\caption{\label{Fig:6} (Color online) Numerical solution of the exchange interaction $J$ scaled by $j_c$ for the electrons (navy curve) and holes (red curve) in monolayer phosphorene for two impurities where the first impurity fixed at the origin and the second one rotated around it on a circle with 5 nm radius by angle $\theta_R$ (see Fig. \ref{fig:lattice}) at finite $\gamma=0.48$ eV nm. The RKKY interaction is strongly anisotropic in phosphorene. The maximum of the susceptibility occurs along the zigzag direction corresponds to $\theta=0,\pi$ and decrease as the direction changes to the armchair direction. Here, the magnitude of the distance vector between two impurities is $R=5$ nm and the Fermi energy is $\varepsilon_{\rm F}=\pm $0.65 eV. }
\end{figure}

Although, we derive our analytic results for two special directions, zigzag and armchair, by using the full numerical relation of the susceptibility, Eq. \eqref{eq:RKKY}, it is possible to obtain the RKKY interaction for any direction in the plane of phosphorene. Figure \ref{Fig:6} shows the dependence of the RKKY interaction with respect to the angle of the distance vector between impurities. This figure clearly shows how this interaction is strongly anisotropic in phosphorene. In this figure, we fix the magnitude of the distance vector between two impurities, $R=5$ nm and also the Fermi energy is $|\varepsilon_{\rm F}|=$0.65 eV. As the RKKY interaction has an oscillatory type of $\sin(2k_{\rm F}R)$ and $k_{\rm F}$ changes on the line of a constant energy, the interaction oscillates with respect to the angle of the distance vector. The maximum of the susceptibility occurs along the zigzag direction corresponds to $\theta=0,\pi$ and decreases as the direction changes to the armchair direction.

\begin{figure}[]
\begin{center}
\includegraphics[width=3.4in]{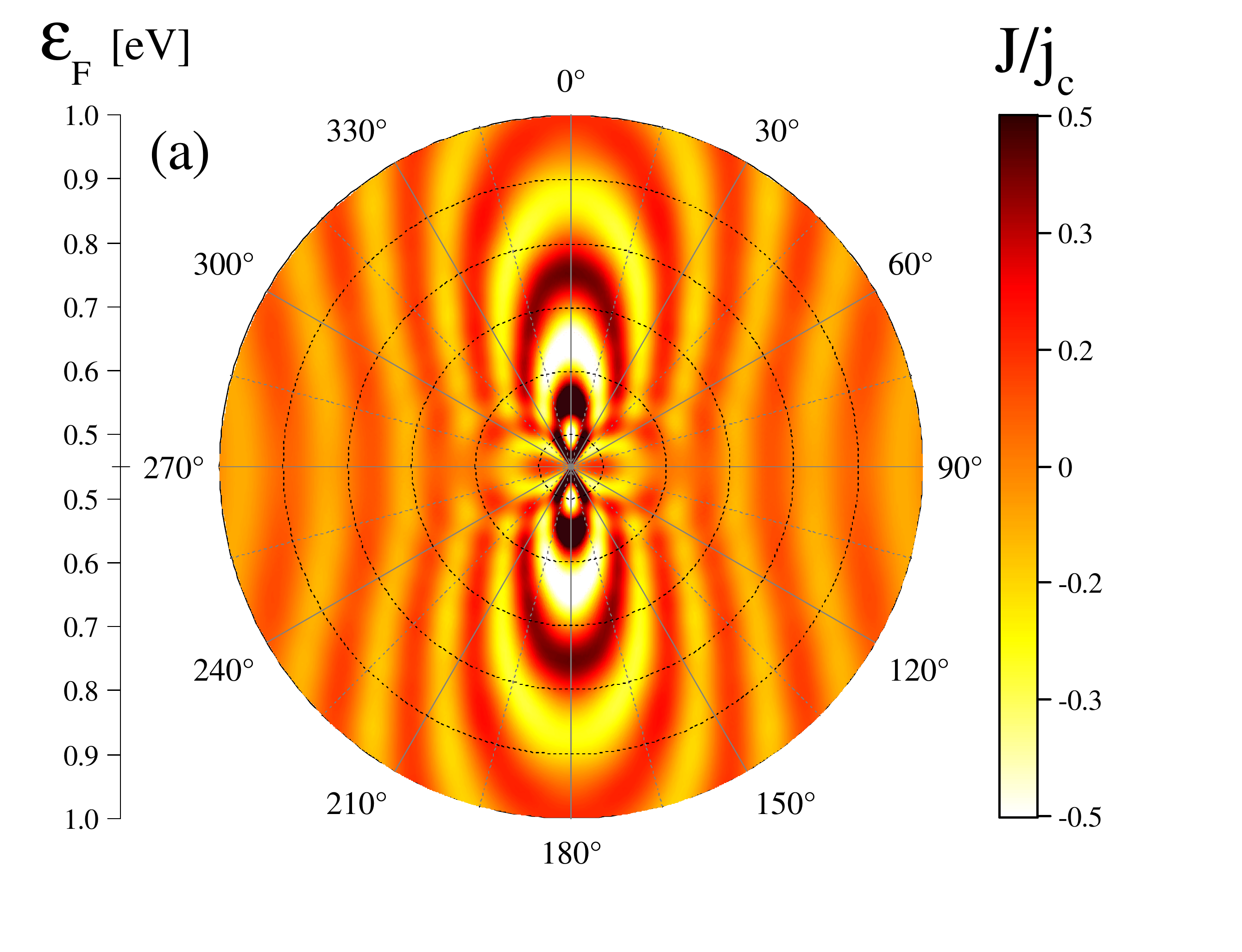}
\includegraphics[width=3.4in]{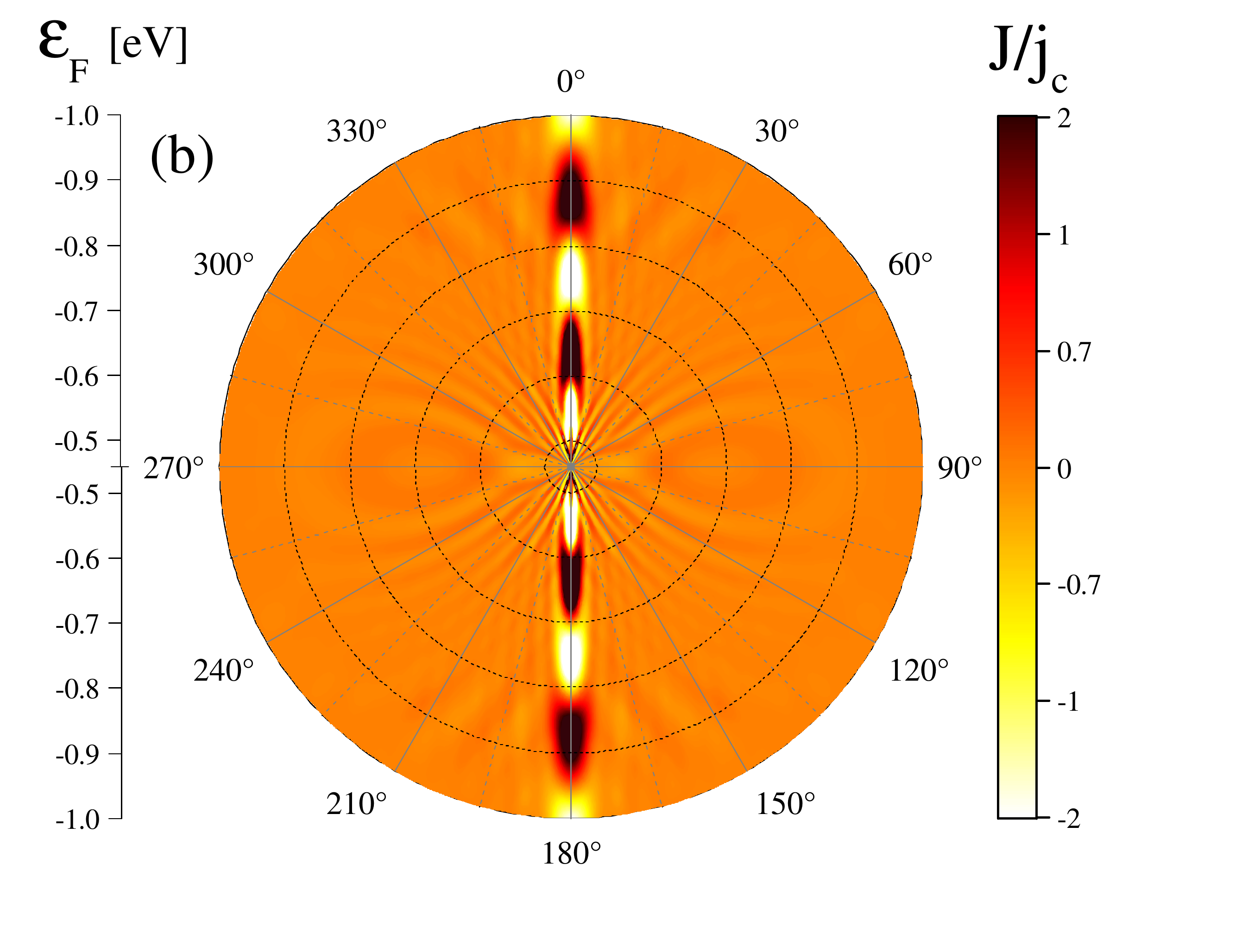}
\end{center}
\caption{\label{Fig:7} (Color online) Normalized RKKY interaction ($J/j_c$) of (a) electron and (b) hole doped system versus the orientation of the second magnetic adatom $\theta_R$, and the Fermi energy of phosphorene $\varepsilon_{\rm F}$ for a fixed distance between two impurities, $R=5$ nm. $\theta_R$ is the angle between the $x$ axis and the axis given by the two impurities (see Fig. 2). A strong anisotropy in a hole doped phosphorene in comparison with the electron doped case is observed. Each circle refers to a value of the Fermi energy.}
\end{figure}

Finally, in Fig. \ref{Fig:7}, we illustrate contour plots of the normalized RKKY interaction ($J/j_c$) for a fixed $R=5$nm and in the plane of distance vector angle and the Fermi energy. This figure again shows that for most of the Fermi energies the RKKY interaction attains its maximum along the zigzag direction. Moreover, by comparing results in (a) and (b) panels which are for n-doped and p-doped phosphorene respectively, one can see a more strongly anisotropy in hole doped phosphorene in comparison with the electron doped case.

\section{summary and conclusion}\label{sec:summary}
 In this work, we have investigated the RKKY interaction in a monolayer phosphorene using the Green's function technique.
We have explored the spatial behavior of the RKKY interaction as well as its dependence on the Fermi energy and direction of the distance vector between magnetic impurities. For the sake of comparison, an analytic expressions for the RKKY interaction within an approximated model Hamiltonian is obtained and afterwards we have compared our full numerical results with those results obtained analytically.
 Due to the flat band at the edge of the valence band and and correspondingly heavy effective mass of the holes along the $Y-\Gamma$ direction, the band curvature is strongly reduced and the RKKY interaction enhances significantly, which means by small amount of doping one can expect large values of the RKKY interaction.
Our numerical results show that the wavelength of the RKKY oscillations fits very well with $\pi/k^s_{\rm F}$ where $k^s_{\rm F}$ is the wavevector of the electrons at the Fermi level in the $s$ (armchair or zigzag) spatial direction.  We also find that the tail of the RKKY oscillations, just like the conventional 2D semiconductors, falls off as $\sin({2k_{\rm F}^s R)}/R^2$ at large distances.
 In both n- and p-doped samples, the RKKY interaction is highly anisotropic especially in the p-doped regime where the highest value of interaction occurs when two impurities are located along the zigzag direction. The value of the susceptibility tends to a minimum value with changing the direction from the zigzag to the armchair direction.

\section{Acknowledgements}

We thank M. Sherafati for useful discussions. F.P. acknowledges "Nordic institute for theoretical physics (Nordita)" for their hospitality while the
last parts of this paper were preparing.  This work is partially supported by the Iran Science Elites Federation and Iran National Science Foundations.

\appendix
\section{Derivation of the real space Green's function }\label{sec:app}
In order to perform the analytical integrations of the case of $\gamma=0$ shown earlier in this paper, we first evaluate the real space Green's function as
\begin{align}\label{eq:a2}
G_0^{v}({\bf R},{0},E)=\frac{1}{\Omega_{BZ}}\int \frac{ �dk_x dk_y}{E-E_v+\eta_v k_x^2 +\nu_v k_y^2} e^{i{ k_x R_x}}e^{i{ k_y R_y}} \nonumber\\
=\frac{1}{\Omega_{BZ}}\int dk_y e^{i{ k_y R_y}} \int \frac{ dk_x}{E-E_v+\eta_v k_x^2 +\nu_v k_y^2} e^{i{ k_x R_x}}.
\end{align}

This integral can be evaluated step by step as follows

\begin{align}\label{eq:a3}
I=\int_{0}^{\infty} \frac{ dk_x}{E-E_v+\eta_v k_x^2 +\nu_v k_y^2} e^{i{ k_x R_x}}=\frac{1}{\eta_v}\int_{0}^{\infty} \frac{ dk_x}{a+ k_x^2} e^{i{ k_x R_x}},
\end{align}
where $a=\frac{E-E_v+\nu_v k_y^2}{\eta_v}$. By using the residues calculations, where\\
 (1) $f(z)$ is an analytic function in the upper half plane except for a finite number of poles.\\
 (2) $\lim_{|z|\rightarrow \infty} f(z)=0$\\
then
\begin{align}\label{eq:a4}
\int_{-\infty}^{\infty} f(x) e^{ia x} dx=2\pi i \sum{residues(upper \ half-plane)}  (a>0).
\end{align}

Here, $f(z)=\frac{e^{iR_x z}}{a+z^2}(k_x\equiv z)$ has two simple poles ($z_{1,2}=\pm i\sqrt{a}$) for which only the pole at ($z_1=+i\sqrt{a}$) lies inside the contour (upper half-plane) thus
\begin{align}\label{eq:a5}
\int_{-\infty}^{\infty} \frac{ dx}{a+ x^2} e^{i{ x R_x}}=2\pi i [f(z)(z-z_1)]_{z=z_1}=2\pi i [\frac{e^{i{R_x z}}}{(z-z_2)}]_{z=z_1}.
\end{align}

After substituting this integration into Eq.~\eqref{eq:a3}, the $I$ is given by
\begin{align}\label{eq:Fourier}
I=\frac{\pi e^{-{R_x \sqrt{a}}}}{2\eta_v \sqrt{a}}=\frac{\pi e^{-{R_x \sqrt{\frac{E-E_v+\nu_v k_y^2}{\eta_v}}}}}{2\eta_v\sqrt{\frac{E-E_v+\nu_v k_y^2}{\eta_v}}}.
\end{align}

\end{document}